\documentstyle[12pt,aaspp4]{article}
\def\ie{i.e.\ }
\def\eg{e.g.,\ }
\def\etal{et~al.\ }
\def\ltsima{$\; \buildrel < \over \sim \;$}
\def\simlt{\lower.5ex\hbox{\ltsima}}
\def\gtsima{$\; \buildrel > \over \sim \;$}
\def\simgt{\lower.5ex\hbox{\gtsima}}
\def\ha{H$\alpha$\ }
\def\fp{Fabry-P\'erot}
\def\kms{km s$^{-1}$}
\def\am{AM~0738--692}

\journalid{}{}
\articleid{}{}
\lefthead{Mihos \& Bothun}
\righthead{NGC 2442}
\slugcomment{To appear in the {\it Astrophysical Journal}}

\begin{document}

\title{NGC 2442: Tidal Encounters and the Evolution of Spiral Galaxies}

\author{J. Christopher Mihos\altaffilmark{1,2}}
\affil{Department of Physics and Astronomy,\break
	 Johns Hopkins University, Baltimore, MD 21218\break
	 hos@pha.jhu.edu}
\and
\author{Gregory D. Bothun}
\affil{Department of Physics,\break
         University of Oregon, Eugene, OR 97403\break
         nuts@moo.uoregon.edu}

\altaffiltext{1}{Hubble Fellow}
\altaffiltext{2}{Visiting Astronomer, Cerro Tololo Inter-American Observatory.
CTIO is operated by AURA, Inc. under contracts to the National Science
Foundation.}

\begin{abstract}

Using imaging \fp\ data, we study the star-forming properties and kinematics
of the nearby barred spiral galaxy NGC 2442. The \ha emission is very
localized along the strong spiral arms of the galaxy, and shows a marked
asymmetry between the sharp, well-defined northern tidal arm and the weaker
southern arm. The \ha velocity field appears highly distorted, with
a rapidly rotating nuclear component. We find evidence for strong 
non-circular motions along the northern arm, coincident with the pronounced 
dust lane and regions of intense star formation. The strong asymmetries, 
disturbed velocity field, and presence of a perturbed companion suggest that 
we are witnessing a strong kinematic response to a close interaction, 
which has redistributed the star formation activity throughout the disk 
of NGC 2442.  Dynamical modeling
of the NGC 2442 system supports this hypothesis, and suggests that the
regions of strongest star formation are coincident with strong shocks
occurring along the tidally perturbed northern arm.   Despite this
strong redistribution of the gas on small scales, this galaxy does not
show a significant departure from the Tully-Fisher relation, nor does it
appear to be experiencing any strong starburst.  Moreover, our models 
predict that in a few x 10$^{8}$ years, NGC 2442 will have recovered from 
this first tidal encounter and will experience another passage  -- and
ultimately a merger -- in a few Gyr. This merger may provoke stronger, 
permanent changes in the structural properties of the galaxy, depending 
on the detailed response of the disk.
Given the environment of many disk galaxies, this tidal encounter 
cycle seems likely to be a normal phase of disk galaxy evolution.

\end{abstract}


\vfil\eject

\section{Introduction}

There is compelling morphological and photometric evidence that enhanced
star-forming activity in disk galaxies is a direct result of tidal interactions
between galaxies.  However, the physics of this strong causal relationship
remain poorly understood.  Although the strongest starbursts are generally 
found in
the nuclei of interacting galaxies, some systems only show star
formation in their outer disks, while others show no sign of elevated
star formation rates at all.   This variety suggests the local physics
of gas dynamical response to perturbations varies considerably from
interaction to interaction.  While models of interaction-induced star formation 
can successfully explain strong central starbursts through bar-induced 
inflows (\eg Noguchi 1988; Mihos, Richstone, \& Bothun 1992;
Mihos \& Hernquist 1994, 1996), they have more difficulty exciting
strong disk star formation. Moreover, induced star formation in interacting
disks is often asymmetric and patchy in nature, suggesting that the 
{\it local} dynamical conditions in the galaxy are an important
piece in the puzzle. To understand in detail how interactions influence
star forming activity, we must simultaneously examine both the kinematics and 
the star forming properties of individual interacting systems.

The southern galaxy NGC 2442, shown in Figure 1, is a nearby (D=16 Mpc
for $H_0=75$ km/s/Mpc) peculiar barred spiral (SAB(s)bc, de Vaucouleurs \etal 
1991 [RC3]) showing many signs indicative of an interaction. Along with 
the short central bar, ovally distorted body, and isophotal twists, 
the disturbed, asymmetric, and extended arms of the galaxy argue for
a collisional origin.  Table 1 summarizes the environment of NGC 2442.
The galaxy has two nearby companions  and is possibly
a member of a loose group (Tully 1988). Because of its proximity,
NGC 2442 offers a rare opportunity to investigate the kinematic
and star forming properties of an interacting galaxy shortly after the
initial collision. At 16 Mpc, 1\arcsec\ subtends 77 pc, affording a very
detailed view into the response of a galaxy to a collisional encounter.

While a complete understanding of the dynamics of interacting galaxies
is impossible when only morphological data exist, the combination
of morphology and kinematic information can break the ambiguities
arising from projection effects and yield a more accurate description
of the dynamical state. Such an approach has been applied previously using 
kinematic information from slit spectra or HI data (\eg Borne 1988;
Borne \etal 1988; Balcells \etal 1989; Stanford \& Balcells 1991; Hibbard \& Mihos 
1995). However, slit spectra do not provide full two-dimensional mapping 
of the ionized gas, while HI observations suffer from relatively low 
spatial resolution. The use of an imaging \fp\ spectrograph
overcomes both these obstacles, providing two dimensional intensity
and velocity maps of the ionized gas at $\sim$ 1\arcsec\ resolution
(\eg Tilanus \& Allen 1991; Vogel \etal 1993, Canzian \etal 1993;
Mihos, Bothun, \& Richstone 1993).  Using the morphology and velocity 
field in conjunction with numerical modeling, it is possible to 
reconstruct the dynamical history of interacting galaxies (\eg Mihos \etal 
1993) and address questions pertaining to the physical triggering mechanism 
for star-forming activity.

Numerical models of colliding galaxies demonstrate clearly that
strong interactions drive rapid dissipation and inflow in the
gaseous components of galaxies (\eg Barnes \& Hernquist 1991, 1996;
Mihos, Richstone, \& Bothun 1992; Mihos \& Hernquist 1994, 1996). 
This inflow is associated with the tidally-induced arms and bars
in the galaxies, as gas is compressed along these features and
driven inwards. In many galaxies, the radial motion along spiral
features easily detectable: $\sim$ 30 \kms\ for M81 (Visser 1980),
$\sim$ 50 -- 70 \kms\ for UGC 2885 (Canzian \etal 1993), $\sim$
60 -- 90 \kms\ for M51 (Vogel \etal 1988, Tilanus \& Allen 1991). 
The fact that M51, closely interacting with NGC 5195, shows the highest 
radial gas motions lends credence to the theoretical models of
strong inflows in galaxy interactions and suggests such motions should
be easily detectable in other interacting galaxies given the resolution
of \fp\ data. The velocity field 
of the ionized gas in and around tidal features should therefore yield 
a great deal of information about the strength of the inflow and the 
physical conditions in the gas.

In this paper, we present imaging \fp\ data on NGC 2442 in order to 
explore the dynamical link between interactions, gaseous
inflow, and star-forming activity. We find evidence for strong shocks and
inflow along the galaxy's northern arm, accompanied by intense, localized
star formation. Very little star formation is occurring in the southern
tidal debris or along the bar itself, and in the inner kpc we detect a 
small, rapidly rotating ring of ionized gas. Based on these observations,
we suggest a scenario in which NGC 2442  has recently (\ie within a few
hundred Myr) experienced a relatively strong interaction with \am, driving the
formation of the tidal arms. Dynamical modeling supports this scenario,
and identifies the strong star forming sites along the arms as regions
experiencing strong shocks and dissipation. The asymmetry observed in
the star forming morphology of the galaxy seems to be related to 
the local dynamical conditions in the evolving tidal debris. 

\section{Observations and Data Reduction}

NGC 2442 was observed in \ha\ with the Rutgers Imaging \fp\ on the CTIO
1.5-m telescope on the night of 1993 April 17. The \fp\ etalons
have a bandpass of $\sim$ 2\AA\ FWHM, giving an instrumental velocity 
dispersion of $\sim$ 50 \kms. To isolate the \ha\ transmission order,
a narrow band ($\sim$ 80\AA) \ha\ filter was used. With the Tek \#4
CCD, the pixel scale was 1.09\arcsec\ pixel$^{-1}$ and the circular field
of view covered $\sim$ 7\arcmin\ in diameter --- enough to contain NGC 2442 
itself, but none of its possible companions. NGC 2442 was scanned in 1\AA\
steps from 6584\AA\ -- 6604\AA, covering the full velocity width of the system.
Observing conditions were not optimal, with seeing varying from 1.5\arcsec\
to 2\arcsec\ over the night and patchy cirrus present on occasion.

After standard bias-subtraction and flat fielding, the individual spectral
images were shifted to a common centroid using foreground stars in the field. 
Sky subtraction proved more problematic, due to night sky emission lines 
in the spectral region scanned.
Because of the radial spectral dispersion of the \fp, these night sky lines
appear in the reddest nine frames as rings of diffuse emission, and had to be 
removed by subtracting model emission lines from each image. Once these sky
lines were removed, a constant sky level could be obtained and subtracted
from each frame.  To account for changes in sky transparency, each image was 
normalized to a common transparency by performing photometry on the stars 
in the field and requiring those stars to have a flat continuum over the 
20\AA\ coverage of the observations. These normalization factors varied from 
5 to 20\%, with typical uncertainties in the range of 5-10\%.
To calibrate the wavelength solution of the instrument, neon arc lamp
exposures were taken several times during the night. 

After the data processing is complete, each pixel in the image has a 20
point spectrum associated with it, from which four pieces of data can
be extracted: the continuum intensity, the integrated \ha\ flux, the 
radial velocity, and the velocity dispersion. These values are obtained
by fitting a Voigt profile to the emission line profiles, binning
pixels in regions of low flux to improve the signal and quality of
fit. Typical errors are \simlt 10 \kms\ in velocity and \simlt 10\%
in \ha\ flux. Because of our relatively short exposure times in each band, 
the signal-to-noise level in the continuum is rather poor. To obtain
a higher quality continuum image, a broad band V image was obtained
by Bob Schommer using the 0.9m telescope at CTIO on 1996 May 14. 
The reduced V-band image, consisting of 4 coadded 10 minute exposures, is shown
in Figure 2.

\section{Morphology and Star Formation}

Figure 3a and 3b show the reduced continuum and \ha\ intensity map for
NGC 2442. The continuum maps (Figures 2 and 3a) reveal a number of NGC 
2442's peculiarities. The bright, compact nucleus is clearly offset from 
the centroid of the outer isophotes by $\sim$ 20\arcsec\ ($\sim$ 1.5 kpc), 
suggesting a strong,
differential distortion of the outer portions of the galaxy. The galaxy
also shows severe isophotal twists -- the major axis of the outer 
isophotes is aligned along a PA of 30\arcdeg, while the major axis of the 
central regions twists from 40\arcdeg\ at 5\arcsec\ to 60\arcdeg\ at 20\arcsec
(Sersic \& Donzelli 1993).  The spiral arms are
quite asymmetric: the northern arm is linear and well defined, extending 
$>$ 20 kpc from the main body, while the curved southern arm is much
more diffuse and shorter, only $\sim$ 13 kpc along its spine. The northern
arm has a strong dust lane traced out over a large portion of the arm,
whereas in the main body and southern arm the dust is more patchy in nature.

The star forming morphology also shows the asymmetry in the spiral arms.
Table 2 summarizes the \ha\ photometry for NGC 2442.\footnote{Note that, because
of the non-photometric observing conditions, we have derived only
relative \ha\ fluxes across the galaxy, and used the calibration by
Dopita \& Ryder (1994) for the photometric zeropoint.}  The bulk of the 
star formation in NGC 2442 is associated with the northern arm, and to 
a lesser extent with the inner portion of the southern arm. Star formation
along these arms is very localized in giant HII regions, with little 
diffuse \ha\ observed. The differences in the star forming activity
between the two arms are most evident in the {\it outer} portion of the
arms. Using the spot at which each arm connects to the main body
of the galaxy (roughly along a line of 30\arcdeg\ PA) to divide the
arms into ``inner'' and ``outer'' regions, we find that the total
amount of star formation in the inner regions of each arm is comparable,
with each arm containing $\sim$ 22\% of the total \ha\ flux from the
galaxy. However, the outer portion of the arms are drastically
different, with the outer northern arm providing 37\% of the total
\ha\ flux of the galaxy, compared to only 9\% from the outer southern
arm. 

The nuclear regions show evidence for both emission from a central point
source as well as a star-forming ring of radius $\sim$ 8\arcsec\ (620 pc;
see Figure 4).
Studies into the nature of the central source in NGC 2442 have been 
inconclusive; using optical spectroscopy, Shobbrock (1966) classified it 
as as Seyfert, while Veron-Cetty \& Veron (1986) call it ``Seyfert-like,''
but showing H$\beta$ in absorption. The ring$+$point-source morphology 
of the \ha\ emission shown here suggests that the confusion is probably
due to the fact that there is a combination of Seyfert (the point
source) and starburst activity (the star-forming ring) occurring in the
nucleus of NGC 2442. The lack of a strong HII region spectrum combined
with the presence of H$\beta$ in absorption suggests that the current
starburst event in the central regions is waning.

Interestingly, although the spatial distribution of \ha\ appears to
have been strongly altered by the interaction, the total \ha\ luminosity
of NGC 2442 ($\log L_{H\alpha}=41.5$; Ryder \& Dopita 1994) is quite normal 
for a non-interacting, luminous, late-type spiral (see \eg Kennicutt \etal
1987). Similarly, NGC 2442 is not infrared luminous, with $\log L_{IR} =
10.0$, and its infrared-to-blue luminosity ratio of $L_{IR}/L_B=0.56$
is typical of isolated spirals and an order of magnitude below interacting
starburst galaxies (see Sanders \& Mirabel 1997). Following Kennicutt,
Tamblyn, amd Congdon (1994), we can estimate the ratio of the current
SFR to the average SFR over the age of the disk using the \ha\ 
and optical (V) luminosity of the disk; for NGC 2442 we derive
$b\equiv SFR/<SFR>_{past}=0.72$, similar to the median value of $<b>=0.85$
for Sbc galaxies derived by Kennicutt \etal (1994). In short,
while the galaxy has responded very strongly in its morphology, kinematics, 
and spatial distribution of star formation, its {\it overall rate} of star 
formation seems quite normal for its Hubble type. Evidently the local 
disturbances spawned by the interaction are more dramatic than the global 
response.

Finally, we point out a small, detached star forming region midway between
the main body and the extended tip of the northern arm. This object,
approximately 500 x 300 pc in size, has an \ha\ luminosity of 
$L_{H\alpha} = 1\times 10^{39}$ erg s$^{-1}$, implying a star formation
rate of $\sim 0.01$ M$_{\sun}$ yr$^{-1}$, using the $L_{H\alpha}$/SFR
conversion factor from Kennicutt (1983). It shows no sign of rotation
in the \ha\ velocity map; however, its systemic velocity (1595 \kms)
is intermediate to the velocity of the outer northern arm and the inner
disk. It is unclear whether this object is a preexisting dwarf companion
to NGC 2442, a massive HII region stripped from the inner disk, or new
star forming region condensing from the tidally perturbed interarm ISM.
However, its small size and \ha\ luminosity ($\sim$ 20\% that of the
Small Magellenic Cloud (Kennicutt \etal 1995)) argues against its status
as a bona-fide dwarf companion (preexisting {\it or} newly formed).
The relatively short lifetime of individual HII regions (\ltsima $10^7$ years;
von Hippel \& Bothun 1990) also make it unlikely to be a stripped disk HII 
region. Models by Barnes \& Hernquist (1992) have demonstrated that
adiabatically cooling gas in tidal debris can collapse and fragment into
clumps. The region between the disk and tidal debris would be a suitable
site for this process to occur, leading perhaps to star forming clumps such
as this.

\section{\ha Kinematics}

Figure 3c and 3d show the derived \ha\ velocity map and velocity dispersion
map for NGC 2442. The FWHM of the velocity distribution is 400 \kms;
however, due to the strong distortion of the galaxy it is difficult to 
unambiguously assign an inclination and derive the rotation speed of the 
galaxy. Assuming the inner ring is circular, we can use its ellipticity 
($b/a = 0.45\pm0.06$) to derive an inclination \footnote{If the ring is 
intrinsically highly elliptical, this method will overestimate
the inclination of the disk. We note that the inclination we derive is only
somewhat higher than that reported in the literature ($i=24$\arcdeg) from an
analysis of the global axial ratio of the arms (\eg Bajaja \& Martin 1985;
Baumgart \& Peterson 1986). However, because of the obvious distortion in
the outer isophotes, use of the global axial ratio to determine inclination
is very suspect.  Furthermore, the derived low inclination would yield an
anomalously large rotation velocity of $v_{circ} = 490$ \kms. More recently,
Ryder and Dopita determined the inclination from the axial ratio of the 
{\it inner} disk and found $i=60$\arcdeg, similar to the value we derive from 
the starburst ring.} of $i=69\pm4$\arcdeg, and an implied circular 
velocity for the system of $v_{circ} = 225$ \kms. We can also use the velocity 
map of the inner ring and nucleus to
solve for the systemic redshift; the median radial velocity for this
region is $v_{sys}=1475\pm10$ \kms. This systemic velocity is greater by
40 \kms\ than the systemic velocity derived via HI (Bajaja \& Martin 1985)
or CO (Bajaja \etal 1995) measurements; this discrepancy can easily 
be explained by the fact that the cold gas is preferentially found on
the blue-shifted NE side of the galaxy (Bajaja \etal 1995). The
mean redshift of the inner ionized ring is therefore a better
estimate of the systemic velocity of NGC 2442.

Using the derived \ha\ velocity map, we attempt to construct a rotation curve
for NGC 2442 in the manner described by Schommer \etal (1993). Using the
inferred inclination $i$ and major axis PA, we can derive the circular
velocity as a function of radius in concentric annuli, shown in Figure 5.
The derived rotation curves are sensitive to the choice of kinematic
center, which is ill-defined in the case of NGC 2442. Two possibilities
are examined: first, that the kinematic center is marked by the nucleus,
and second, that the kinematic center is defined by the centroid of the
disk isophotes. As noted previously, the nucleus is offset from the
isophotal center of the galaxy by 20\arcsec, and because of the lack of 
\ha\ emission from R=10\arcsec--70\arcsec, it is difficult to discern 
kinematically which option is more physically reasonable. With 
nucleus-centering, the rotation curve shows a rapid rise, then falls
with radius; with isophotal-centering, the rotation curve is rising
over much of the disk. Further complicating the rotation curve derivation
is the fact that the \ha\ emission is preferentially found along the
tidally perturbed spiral arms, where non-circular motions should be highest. 
Tests using the numerical models described in \S 6 indicate that the shape 
of the derived rotation curve is sensitive to the azimuthal portion of the 
disk sampled by the \ha\ emission. This kinematic bias,
coupled with the sparseness of the \ha\ data and the uncertainty in the 
kinematic center, does not allow for a physically meaningful constraint
on either the mass distribution or the global kinematics of NGC 2442;
as a result we focus on the two-dimensional
velocity map rather than the rotation curve in our discussion of
kinematics and model matching below. 

One thing {\it is} clear from the rotation curve, however:
the ionized gas in the nuclear region shows a steep rise in rotation 
velocity, reaching a maximum of $v\sin i = 220\pm10$ \kms\ only 8\arcsec\
(500 pc) from the nucleus. Unlike the outer body, the velocity field of the
inner ring does show a simple spider-diagram indicating circular motion
and we derive a mass interior to the ring of 
$M_{nuc}=Rv^2/G = 8.4\pm1.5\times 10^9$M$_{\sun}$. The rapidly rising rotation
curve and large mass interior to 500 pc is consistent with the bright,
compact nucleus seen in the continuum map. The velocity dispersion of
the ionized gas in the ring is rather high, $\sim$ 50--60 \kms, while
in the central source itself the velocity dispersion reaches as high
as 125 \kms. 

Striking evidence for strong noncircular motions in NGC 2442
comes from the kinematic patterns evident in the two-dimensional
velocity map (Figure 3c). In the main body of the galaxy these 
velocity patterns can also be seen
in the CO map of Bajaja \etal (1995), albeit at much lower spatial
resolution. In the classic ``spider diagram'' for a disk in circular
rotation, isovelocity contours fan out in a V-shaped pattern along
the major axis. In such a velocity field, cuts parallel to the minor
axis will show velocity symmetry across the major axis. NGC 2442 shows 
no such behavior; rather, the NW and SE portions of the disk show skewed 
velocity patterns. On the SW side of the disk, a cut parallel to the 
minor axis at R=75\arcsec\ (5 kpc), shows a rather large velocity difference of
$\sim$ 120 \kms\ between the NW and SE extrema of the cut. A similar
cut along the NE side of the disk shows a velocity difference of $\sim$
150 \kms\ across the cut. In essence, rather than being symmetric around
the minor axis (PA = 120\arcdeg), the isovelocity contours in the inner disk
run nearly N-S, along a PA of 0\arcdeg. The fact that the isovelocity
contours are distorted along the spiral arms is a clear indication of 
strong non-circular motion induced by the arms. Further out in the tidal 
arms, the isovelocity contours twist again, following a PA of 155\arcdeg. 

The most curious portion of the velocity field lies in the NE side
of the disk, where the northern tidal arm connects to the galaxy. The very 
strong
dust lane cuts through this region, and the strongest sites of star formation
are located here as well. Running parallel to the dust lane, but offset to
the inside by 4\arcsec\ (250 pc), is a low velocity ``trough'' in the velocity
field of the arm (see Figure 6). In this portion of the arm, the isovelocity
contours run {\it parallel} to the arm, dropping from 1280 -- 1290 \kms\ on
the inside of the arm to 1230 -- 1240 \kms\ in the trough, before rising 
again to 1280 -- 1290 \kms\ on the outside of the arm. Along this trough,
the velocity dispersion in the gas reaches its highest (non-nuclear) values:
75 -- 85 \kms, compared to more typical values of 30 -- 40 \kms\ in
other regions of the galaxy and tidal arms. The velocity dispersions
are increased from the point where the arm enters the disk, and continue
along the arm until the point where the arm turns back towards the inner 
regions of the disk.  The strongest sites of \ha\ emission occur on the 
{\it outside} of the dust lane; however, it is unclear whether this reflects an 
offset in the physical location of the star formation, or merely strong 
obscuration of the \ha\ line in the dust lane.

Finally we note the velocity structure along the outer portion of the northern
arm. This tidal arm shows a modest velocity gradient along the spine of the
arm, such that regions further along the tail are subsequently more redshifted
(by 200 \kms\ over the 20 kpc span of the arm).  However, the isovelocity
contours run largely parallel to the tidal arm, in the sense of a smooth
gradient (rather than velocity trough) across the arm. Perpendicular cuts
through the arm reveal velocity differences of $\sim$ 100 \kms\ across the
kiloparsec breadth of the arm, similar to the gradients observed in the
tidal arms of the interacting galaxy NGC 6872 (Mihos \etal 1993). 

The combined information in the morphology and velocity maps presented in
Figures 2, 3, and 6 suggests the following scenario along the northern edge
of NGC 2442's disk. The tidal interaction has drawn out the northern arm
from the main body of the galaxy, compressing the gas along the spine
of the arm. As shown by numerical modeling of tidal features (\eg
Hernquist \& Spergel 1992; Hibbard \& Mihos 1995; see also \S 6 below),
material in the tidal arm now falls back inwards towards the main body. As 
gas returns to the disk, it is shocked and compressed where the arm and disk 
intersect. A similar fate may hold for gas on the W side of the disk, rotating 
into this contact point.  The high velocity dispersion where gas
enters the shock (upstream from the dust lane) attests to the turbulent
motion in the preshock region. As the gas is compressed along the shock,
it dissipates energy and begins to flow inward, decoupling from the motion
of the surrounding arm (and thus producing the velocity trough). Downstream
from the shock, star formation occurs in the cooling postshock gas, as 
evidenced by the strong \ha\ emission along the outside ridge of the dust 
lane.

While these local effects are dramatic it is important to verify if
interactions like these are so severe to nullify including this galaxy
in any kind of Tully-Fisher (TF) analysis.  Recall that an imaging study of
this galaxy might not necessarily flag it as being strongly interacting
as there are currently no obvious close, luminous companions.   We can
place NGC 2442 on the TF relation by using its observed R magnitude of
$R=9.71 \pm 0.09$ (RC3) and a distance modulus of 31.0 $\pm$ 0.3 (Sersic
\& Donzelli 1993), and correcting for galactic absorption via Burstein
\& Heiles (1984) and internal absorption via Tully \& Fouque (1985),
after which we derive
an absolute R magnitude of $M_R=-22.1 \pm 0.3$. Using the HI velocity
width from Bajaja \& Martin (1985), and correcting for inclination, we can
now put NGC 2442 on the R-band TF relationship derived by Pierce
\& Tully (1992). Despite the uncertainties in the inclination and internal
absorption, Figure 7 shows that NGC 2442 lies essentially on the TF
relation defined by the local calibrators. Although the detailed
kinematics of the galaxy have been significantly affected by the interaction,
surprisingly, its overall dynamical behavior has not. Our results suggest that
studies of the TF relation at higher redshift will not be compromised
by effects due to galaxy interactions, save for specific cases
where the interaction is very strong (\ie a merger).

\section{Who's to Blame?}

The disturbed morphology and velocity field of NGC 2442 suggest strongly
that the galaxy was involved an encounter in the not-too-distant past.
NGC 2442 may be a member of a loose group (Tully 1988), and several
nearby neighbors do exist; in particular, the E0 galaxy NGC 2434
and the small SB0/a \am. Both galaxies lie at the same
redshift as NGC 2442 (see Table 1).  \am\ is the closer of the two companions,
at a projected separation of 50 kpc; however, it is fainter
by 2.2 magnitudes in R than NGC 2442, suggesting a rather large mass 
ratio (assuming similar M/L's for the galaxies) of $\sim$ 7.5:1.
Although twice as far away as \am, NGC 2434 is brighter by almost 
a magnitude in R, and the implied mass ratio between NGC 2442 and NGC 2434 
is more hefty 3:1. No other prominent galaxies lie with a projected 
distance of 400 kpc of NGC 2442 (Table 1).

Despite its low mass, we favor \am\ as the perturber due to its
disturbed morphology. Figure 8 shows the morphology of \am\
taken from the Digital Sky Survey. The galaxy has a very asymmetric
light distribution, with highly twisted outer isophotes. In fact,
the outer arms look suspiciously like the broad tidal tails
which would develop in a galaxy with large $\sigma/v$. A spectrum of
\am\ reveals a pure stellar component (\ie no emission lines indicative
of star formation)
with a velocity dispersion of of $\sigma \sim 120$ \kms\ (G. Aldering,
private communication), consistent with the SB0/a classification of the 
galaxy, and the lack of strong tidal arms normally spawned in interactions
involving cold stellar disks. Because of the relatively high inferred mass 
ratio for the galaxy pair, the collision must have happened quite 
recently -- within the past few galactic rotation periods -- for NGC 2442 to 
still show such strong asymmetries and velocity distortions. 

\section{Dynamical Modeling}

In order to gain more insight into the dynamical history of NGC 2442,
we simulate a NGC 2442-type encounter, following the evolution of
both the stellar and gaseous components of the galaxies. Recent attempts
to model specific interacting systems have yielded reasonable success
(see, \eg Barnes 1988; Mihos \etal 1993; Hibbard \& Mihos 1995;
Hibbard \& Barnes 1996), although the derived solutions
depend critically on the chosen dark matter
distribution in the galaxies (Mihos, Dubinski, \& Hernquist 1996). This 
result suggests that any given model may suffer from a severe 
lack of uniqueness due to the poorly constrained dark matter distribution.  
Rather than attempting an exhaustive survey of orbital 
configurations and galaxy models to recreate the system in minute detail, we 
choose to focus on an encounter which captures the important features of 
the NGC 2442 system. 

The initial galaxy models are constructed using the technique described
by Hernquist (1993), wherein the primary galaxy consists of a spherical
dark halo, an exponential disk of both stars and gas, and a compact
central bulge. We use a system of units in which the gravitational
constant $G=1$, the disk scale length $h=1$, and the disk mass $M_d=1$.
The disk is populated with both star and gas particles out to a maximum
radius of $R_d=15$.  The halo is modeled as a ``truncated" isothermal 
sphere of mass $M_h=5.8$, core radius $\gamma=1$, exponential cutoff 
radius $r_t=10$, and maximum radius $R_h$=30. The central bulge follows a 
flattened Hernquist (1990) profile, with scale length $a=0.14$, flattening 
$c/a=0.5$ and total mass $M_b=1/9$. The resulting rotation curve for the 
model is similar to the ``universal'' rotation curve for bright spirals 
defined by Persic, Salucci, \& Stel (1996).  The gas comprises 10\% of the
disk mass, and is distributed in the same exponential profile defined by the
disk stars. For expediency, an isothermal equation of state is chosen for
the gas, with temperature $T_{gas}=10^4$ K. The companion galaxy is modeled 
as a gas-free spherical Hernquist (1990) model with total mass $M_c=0.90$. 
The companion mass was chosen to give a mass ratio between the galaxies 
of 7.5, similar to the observed R band luminosity ratio between NGC 2442
and \am.  For the final models shown
here, a total of 102,400 particles were used: 32,768 each in the stellar
disk and dark halo; 4,096 in the central bulge; 16,384 in the companion
galaxy; and 16,384 SPH particles to follow the evolution of the disk gas.
The simulation was evolved using a combined N-body/smoothed particle
hydrodynamics code (TREESPH; see Hernquist \& Katz 1989), and run on
the Cray C90 at the San Diego Supercomputing Center.

Having selected the galaxy models, the remaining parameters describing
the system are the orbital energy, geometry, collision time, and
viewing angle. We choose a parabolic (zero-energy) orbit for the
encounter; hyperbolic encounters are unable to excite strong tidal
features at this mass ratio, while a tightly bound orbit is unlikely
given the (relatively) large current separation of the galaxy pair.
In order to gain some constraint on the orbital parameters, we ran a set
of pure $N$-body models (\ie no gas), varying perigalactic distance
($R_p$), orbital inclination ($i$), and argument of periapse ($\omega$;
see \eg Toomre \& Toomre 1972). Not surprisingly, degeneracies exist 
between the models when viewed from different angles and at differing 
times. Nonetheless, some constraints were clear. No model could
successfully reproduce the strong coherent NW tidal arm at times much later 
than a rotation period after the initial collision. Regardless of 
orbital inclination, no model was able to torque disk material significantly
out of the disk plane. Finally, very distant encounters had trouble
exciting tidal features, while in close encounters the companion galaxy
remains too close to the primary, when compared to the NGC 2442 system.
Our best ``approximate match" model uses $R_p=5.0,\ i=\omega=45$\arcdeg,
and is observed at 20 time units ($\sim$ 1.5 half-mass rotation periods)
past perigalacticon.

Figure 9 shows the evolution of the NGC 2442 model, observed in the
plane of the sky. Only the stellar disk and companion are shown in
the figure. The galaxies reach perigalacticon at T=25, and the current
viewing time is T=46. To convert from model units to physical units,
we scale the model in size and velocity such that the disk size (measured
along the major axis PA from arm to arm) and FWHM of the velocity distribution
match the observed values. With this scaling, unit length is 3 kpc, unit 
velocity is 265 \kms, unit mass is $4.8\times 10^{10}$ M$_{\sun}$, and 
unit time is 11.5 Myr. The scaled
time since collision is thus 230 Myr; at this time, the galaxies have 
reached a projected separation of 48 kpc and the companion has line-of-sight
velocity of +25 \kms\ (with respect to NGC 2442), comparable to the observed 
values of 50 kpc and +54$\pm53$ \kms. The inclination of the galaxy to the sky 
plane is $\sim$ 60\arcdeg. At the observed time, the model is not a
perfect match -- in particular, the pitch angle of the NW arm is smaller 
in the model, and the strength of the southern arm is somewhat stronger. 
Nonetheless, these discrepancies are fairly minor, and this simulation should 
provide a reasonable description of the dynamical state of the NGC2442 system.

Even at this relatively large mass 
ratio, the encounter has no trouble exciting strong tidal features like
those observed in NGC 2442, yet the companion still has managed to reach
the necessary separation of 50 kpc. As the companion passes by the primary,
it draws material out in a rather diffuse, illusory bridge.
In particular, note time T=40, where this ``bridge" consists of
a spray of material drawn from the nearside of the disk, while the strong
tail on the opposite side is more coherent in structure, flaring only near 
its end. As the model evolves towards the current time, material falls
back towards the disk from both tidal features, compressing along the
spiral features in the disk. The gas densities are highest along the
spiral arms and in the nuclear region (see Figure 10). In addition,
the NW tidal arm is well separated from the main disk, while the
southern tidal debris is more diffuse, with a continuous spray of
material from the disk out to the outer edge of the SE arm. 

With the kinematic information provided in the model, we can look at the
physical conditions in the gas along the tidal arm, the analogue of 
NGC 2442's strong NW arm. Figure 11 shows the velocity structure in the
model, with vectors showing the particles' projected tangential velocity 
in each frame. The strong streaming along the spiral arms is apparent; as 
material enters the arms, it shocks and begins to flow inwards along the arm.
In projected radial velocity, the same skewed velocity field is seen in
the model as in NGC 2442. Inside the inner scale length of the disk,
the arms weaken in strength due to the stabilizing influence of the bulge
and the velocity patterns become more circular; however, some net inflow
still occurs.

Focusing on the kinematic structure in the NW arm, we see that the
point at which the arm connects into the main body is in fact dominated
by orbit crossing in the gas, as material from the disk streams into
the tidal arm and dissipates energy. In NGC 2442, this point is characterized
by strong star formation and the low velocity trough. The trough and the
streaming along the outer arm suggested in the observed velocity of NGC 2442
are visible in a line-of-sight radial velocity map of the model, but to a 
lesser degree
than observed in NGC 2442. Most likely this is a resolution effect -- while
the global kinematic structure of the tidal arm is well reproduced, 
the hydrodynamic smoothing length of the gas particles is comparable to the
width of the arm, making substructure within the arm hard to resolve.
Alternatively, there is some indication that the tidal perturbation on
the galaxy is even greater than suggested by the model -- the offset nucleus,
the wide pitch angle of the NW arm -- which may result in a stronger
decoupling of gas velocities along the tidal arm. 

On the whole, our model does a reasonable job of matching both the morphology
and kinematics of the NGC 2442/\am\ pair. In particular, we are able to 
match the asymmetry in 
the tidal arms, the relative position and velocity of \am, and the skewed 
velocity field of NGC 2442. The model also reproduces the kinematic features 
along the northern side of the disk associated with shocks and dissipation in
the gas returning from the NW tidal arm. Some discrepancies remain, however,
including the too-prominent southern arm, the lack of an offset nucleus, 
and the smooth distribution of gas in the inner disk. Further improvements
on the model are possible, and could take the form of differing dark
matter distributions (to alter the disk response), orbit geometries (to
enhance the asymmetry of the tidal arms), and/or gas distributions in
the pre-encounter disk of NGC 2442. Nonetheless, we are encouraged by
the quality of the model, given the minimal amount of ``tuning'' of 
parameters in our model matching effort.

What is the future fate of our modeled encounter, and, by extension, the
NGC 2442/\am\ pair? After another disk rotation period ($\sim$ 350 Myr
after the collision), much of the material currently in the extended tidal 
debris has begun to fall back towards the disk and tidal features are becoming 
quite diffuse.  In the disk, a strong grand-design spiral structure persists, 
but after 500 Myr distinct signatures of the tidal encounter are hard to 
identify. Although warped and somewhat heated by the tidal encounter,
the disk galaxy appears largely axisymmetric, with no strong tidal distortions.
The action is not yet finished, however;
orbital energy has been transferred to internal motions of the galaxies,
which find themselves on a loosely bound orbit. We have followed the
subsequent evolution of the galaxies and find that after reaching a maximum
separation of $\sim$ 100 kpc, \am\ begins to fall back towards NGC 2442,
with a second close passage ($r_{peri} \sim 20$ kpc) occurring in another
2 Gyr. After this, the orbital decay is quite rapid, and following a third
close encounter the galaxies ultimately merge after a total elapsed time 
of 3 Gyr. This merger may be quite destructive to the disk of NGC 2442.
In our model, the companion survives the intermediate passages with most 
of its mass intact, so that it has a strong impact when it reaches the
disk of NGC 2442. As a result, the thin, cold disk is largely destroyed
by the merger. The merged object is still highly flattened, however, and
shows significant rotation. Rather than resembling the very elliptical-like
remnants formed in major mergers of comparable mass galaxies (\eg Barnes
1992; Barnes \& Hernquist 1996; Mihos \& Hernquist 1996), this remnant
may be more akin to S0 galaxies, with large bulge:disk ratios, diffuse
disks, and high $\sigma/v$.

We caution, however, that the final results of the model are strongly
dependent on both the distribution of dark matter around the galaxies
and the internal structure of the companion galaxy. Unfortunately, neither of
these values are well constrained. Because the density and dark matter
content of the galaxies determines the degree to which the companion
is stripped of mass, a lower density companion, coupled with an extended
distribution of dark matter around NGC 2442, could sufficiently strip
the companion of mass, extending the merging timescale and reducing the
impact on the disk of NGC 2442. While the final outcome of such intermediate
mass mergers is crucial to understand, a proper treatment of the problem
is beyond the scope of the models presented here.

\section{Summary}

Using \fp\ mapping of the morphology and kinematics of NGC 2442 in
conjunction with a representative numerical model of the system,
we have explored the triggering of star forming activity by a close
interaction. Our results indicate that in the recent past, $\sim$
150 -- 250 Myr ago, NGC 2442 and \am\ experienced a close encounter, resulting
in the tidal distortions currently observed in both galaxies. The distorted
isophotes and velocity patterns throughout the disk of NGC 2442 attest to the
damage done by even this relatively high mass ratio encounter.
Material on the side of the disk closest to the companion experienced 
significant tidal shear, shredding much of the coherent tidal structure. 
Conversely, material on the far side of the disk was subject to a more 
coherent tidal compression,
resulting in the formation of the dramatic northern tidal arm. As gas
returns to the disk along this arm, it is being shocked and recompressed,
forming a ridge of intense star formation and radial inflow along the
NGC 2442's northern edge.

The asymmetric pattern of star formation in NGC 2442 is a reflection
of the differing physical conditions in the gas on either side of the 
galaxy. On the northern side of the galaxy, the gas is undergoing strong 
shocking and dissipation along the tidally induced arm, resulting in the 
vigorous, localized star formation. The extended tidal arm on the south 
side of the galaxy is much more diffuse, and the gas is not as strongly 
compressed along this feature. This kinematic structure is a reflection
of the formation history of the tidal arms. Material close to the
companion at perigalacticon experiences a very strong tidal shear, dragging 
material out into a transient tidal bridge which rapidly disperses (\eg
Toomre \& Toomre 1972; Mihos \etal 1992, 1993). In contrast, material on 
the opposite side of the disk feels a gentler -- but still effective -- 
tidal field, resulting in a more coherent tidal tail (or, in a high mass 
ratio encounter such as this, tidal arm). In this scenario, the south side 
of NGC 2442's disk was the closest point to the passing companion; in the 
subsequent time since the encounter, the outer disk has rotated significantly,
now presenting the NE side closest to \am in projection. 

The subsequent evolution of the NGC 2442 system is unclear. It is likely
that the interaction will have extracted sufficient energy to bind the
galaxies, and in our model the galaxies will merge in $\sim$ 3 Gyr.  In 
principle, this ultimate merger may be violent enough to lead to a 
transformation of NGC 2442 from Sbc to Sa or even S0, depending on the 
detailed structure of the merging galaxies. In the meantime, NGC 2442 will 
have many rotation periods to resettle into a relatively normal 
configuration.  While the local gas dynamical response to these first 
encounters can be dramatic, the collisions don't seem to significantly alter 
the basic structure and global properties of large disk galaxies, until the 
orbit decays sufficiently to lead to a merger.
We are most probably catching NGC 2442 in a rather transient but common 
phase of disk galaxy evolution -- recovering from a recent collision,
and awaiting another yet to come.  

\acknowledgements

We thank Bob Schommer for obtaining the V band image of NGC 2442, and
Greg Aldering for the spectrum of \am.  We also thank Ted Williams for 
observing support, Ben Weiner and Charles Beauvais for helpful suggestions 
during the data reduction, Stuart Ryder for providing the $H\alpha$ 
flux calibration, and Gerhardt Meurer for a helpful review of the manuscript.
Marc Balcells provided a thorough referee's report which improved the 
presentation of the paper. J.C.M. is supported by NASA through a Hubble 
Fellowship grant \#~HF-01074.01-94A awarded by the Space Telescope Science 
Institute, which is operated by the Association of University for Research in 
Astronomy, Inc., for NASA under contract NAS 5-26555. Partial computing support
for this project was provided by the San Diego Supercomputer Center.

This research has also made use of the NASA/IPAC Extragalactic Database (NED) 
which is operated by the Jet Propulsion Laboratory, California Institute of 
Technology, under contract with the National Aeronautics and Space 
Administration. The Digitized Sky Survey images are based on photographic data 
obtained using The UK Schmidt Telescope.  The UK Schmidt Telescope was 
operated by the Royal Observatory Edinburgh, with funding from the UK Science 
and Engineering Research Council, until 1988 June, and thereafter by the 
Anglo-Australian Observatory.  Original plate material is copyright (c) the 
Royal Observatory Edinburgh and the Anglo-Australian Observatory.  The 
plates were processed into the present compressed digital form with 
their permission.  The Digitized Sky Survey was produced at the Space 
Telescope Science Institute under US Government grant NAG W-2166.

\clearpage

\begin{figure}
\caption{A Digitized Sky Survey image, centered on NGC 2442, also showing
NGC 2434 (upper right) and \am\ (left). The image measures 30\arcmin\ on 
a side. North is up; east is to the left.}
\end{figure}

\begin{figure}
\caption{Broad band V image of NGC 2442. North is up; east is to the left.}
\end{figure}

\begin{figure}
\caption{Reduced \fp\ maps of NGC 2442. a) Continuum map, b) \ha\ flux map,
c) velocity map, b) velocity dispersion map. In the velocity map, light
shading represents blueshift, while dark shading represents redshift (relative
to systemic). In the velocity dispersion map, dark shading indicates high
velocity dispersion.}
\end{figure}

\begin{figure}
\caption{Close up image of the \ha\ intensity map in the inner disk.}
\end{figure}

\begin{figure}
\caption{Rotation curves derived from \ha\ velocity maps. The left panel
shows the rotation curve derived with the nucleus as the kinematic center 
of the galaxy, while the right panel shows the rotation curve derived
assuming the kinematic center is defined by the centroid of the disk
isophotes.}
\end{figure}

\begin{figure}
\caption{Close up image of the \ha\ velocity and velocity dispersion
maps in the northern arm.}
\end{figure}

\begin{figure}
\caption{NGC 2442 on the R-band Tully-Fisher relationship of Pierce \&
Tully (1992). The dashed lines represent the $\pm 1 \sigma$ dispersion in 
the relationship. Errorbars on the NGC 2442 point represent a $\pm 0.3$
magnitude error in the distance modulus and a $\pm 4$\arcdeg error in
inclination.}
\end{figure}

\begin{figure}
\caption{A close-up from the Digitized Sky Survey image, centered 
on \am. Note the distorted fan-like arms.}
\end{figure}

\begin{figure}
\caption{Evolution of the NGC 2442 model, viewed in the ``sky plane." Only
particles from the stellar disk and companion are shown. The frames measure
20 length units on a side, and time is shown in the upper right.}
\end{figure}

\begin{figure}
\caption{Grayscale representation of the gas distribution in the NGC 2442 model
at the best match time.}
\end{figure}

\begin{figure}
\caption{The velocity field of the gas in the NGC 2442 model at the best match
time. The vectors represent the velocity vectors projected onto the plane of
the figure. The left side shows the model observed in the sky plane; the right 
side shows the model observed in the disk plane.}
\end{figure}

\clearpage
 
\begin{table}
\caption{Galaxies in the NGC 2442 Field}
\begin{tabular}{cccc}\hline
Galaxy & $R_T\tablenotemark{a}$ & $cz$ & $R_{proj}$ \\ 
 & & (\kms) & (kpc) \\ \hline
NGC 2442 & 9.71$\pm$ 0.09 & 1475$\pm$10\tablenotemark{b} & -- \\
AM 0738--692 & 11.91$\pm$ 0.09 & 1529$\pm$53\tablenotemark{c} & 50 \\
NGC 2434 & 10.84$\pm$ 0.09 & 1390$\pm$27\tablenotemark{a} & 84 \\
NGC 2397 & 11.57$\pm$ 0.09 & 1363$\pm$10\tablenotemark{d} & 420 \\ \hline
\end{tabular}
\tablenotetext{a}{from the RC3.}
\tablenotetext{b}{this work.}
\tablenotetext{c}{G. Aldering, private communication.}
\tablenotetext{d}{Mathewson, Ford, \& Buchhorn 1992.}
\end{table}
 
\begin{table}
\caption{\ha\ Photometry of NGC 2442}
\begin{tabular}{ccc}\hline
Region & \ha\ Flux\tablenotemark{a} & Fraction of \\
 & ($10^{-13}$ erg s$^{-1}$ cm$^{-2}$) & Total Flux \\ \hline
Outer N arm & 38.7 & 37\% \\
Inner N arm & 23.0 & 22\% \\
Inner S arm & 23.0 & 22\% \\
Outer S arm & 9.4 & 9\% \\
Nucleus & 4.7 & 4.5\% \\
Body & 6.3 & 6\% \\
Detached Clump & 0.31 & 0.3\% \\ \hline
\tablenotetext{a}{using the total flux calibration of Dopita \& Ryder 1994.}
\end{tabular}
\end{table}


\clearpage

\begin{references}

\reference{B1} Bajaja, E., Wielebinski, R., Reuter, H.-P., Harnett, J.I., 
	\& Hummel, E. 1995, \aaps, 114, 147
\reference{BM} Bajaja, E., \& Martin, C.M. 1985, \aj, 90, 1783
\reference{BBH2} Balcells, M., Borne, K.D., Hoessel, J.G. 1989, \apj, 336, 655
\reference{B88} Barnes, J.E. 1988, \apj, 331, 699
\reference{B88} Barnes, J.E. 1992, \apj, 393, 484
\reference{BH91} Barnes, J.E., \& Hernquist, L.E. 1991, \apj, 370, L65
\reference{BH92} Barnes, J.E., \& Hernquist, L.E. 1992, Nature, 360, 715
\reference{BH96} Barnes, J.E., \& Hernquist, L.E. 1996, \apj, 471, 115
\reference{BP} Baumgart, C.W., \& Peterson, C.J. 1986, PASP, 98 56
\reference{B88} Borne, K.D. 1988, \apj, 330, 61
\reference{BBH} Borne, K.D., Balcells, M., \& Hoessel, J.G. 1988, \apj, 333, 567
\reference{BH} Burstein, D., \& Heiles, C. 1984, \apjs, 54, 33
\reference{C1} Canzian, B., Allen, R.J., Tilanus, R.P.J. 1993, \apj, 406, 457
\reference{D1} Dahlem, M., Bomans, D.J., \& Will, J.-M. 1994, \apj, 432, 590
\reference{RC3} de Vaucouleurs, G., de Vaucouleurs, A., Corwin, H.G., 
	Buta, R.J., Paturel, G., \& Fouque, P. 1991, Third Reference Catalogue 
	of Bright Galaxies, Springer-Verlag (RC3)
\reference{DR} Dopita, M.A., \& Ryder, S.D. 1994, \apj, 430, 163
\reference{H90} Hernquist, L. 1990, \apj, 356, 359
\reference{H93} Hernquist, L. 1993, \apjs, 416, 9
\reference{HK} Hernquist, L., \& Katz, N. 1989, \apjs, 70, 419
\reference{HS} Hernquist, L. \& Spergel, D.N. 1992, \apj, 399, L117
\reference{HM} Hibbard, J.E., \& Mihos, J.C. 1995, \aj, 110, 140
\reference{HB} Hibbard, J.E., \& Barnes, J.E. 1996, in preparation
\reference{K2} Kennicutt, R.C. 1983, \apj, 272, 54
\reference{K1} Kennicutt, R.C., Bresolin, F., Bomans, D.J., Bothun, G.D., \&
	Thompson, I.B. 1995, \aj, 109, 594
\reference{K2} Kennicutt, R.C., Tamblyn, P., \& Congdon, C.E. 1994, \apj, 
	435, 22
\reference{MFB} Mathewson, D.S., Ford, V.L., \& Buchhorn, M. 1992, \apjs, 
	81, 413
\reference{MBR} Mihos, J.C., Bothun, G.D., \& Richstone, D.O. 1993, \apj, 418, 
	82
\reference{MDH} Mihos, J.C., Dubinski, J., \& Hernquist, L. 1996, in preparation
\reference{MRB} Mihos, J.C., Richstone, D.O., Bothun, G.D. 1992, \apj, 400, 153
\reference{MH94} Mihos, J.C., \& Hernquist, L. 1994, \apj, 431, L9
\reference{MH96} Mihos, J.C., \& Hernquist, L.E. 1996, \apj, 464, 641
\reference{PSS} Persic, M., Salucci, P., \& Stel, F. 1996, \mnras, 281, 27
\reference{PT} Pierce, M.J., \& Tully, R.B. 1992 \apj, 387, 47
\reference{RD} Ryder, S.D., \& Dopita, M.A. 1994, \apj, 430, 142
\reference{SM} Sanders, D.B., \& Mirabel, I.F. 1997, \araa, in press
\reference{S}  Schommer, R.A., Bothun, G.D., Williams, T.B., \& Mould, J.R. 
	1993, \aj, 105, 97
\reference{SD} S\'ersic, J.L. \& Donzelli, C. 1993, \aaps, 98, 21
\reference{SB} Stanford, S.A., \& Balcells, M. 1991, \apj, 370, 118
\reference{TA} Tilanus, R.P.J., \& Allen, R.J. 1993, \aap, 274, 707
\reference{TT} Toomre, A., \& Toomre, J. 1972, \apj, 178, 623
\reference{TF} Tully, R.B., \& Fouque, P. 1985, \apjs, 58, 67
\reference{VV} V\'eron-Cetty, M.-P., \& V\'eron, P. 1986, \aaps, 66, 335
\reference{V1} Visser, H.C.D. 1980, \aap, 88, 149
\reference{V2} Vogel, S.N., Rand, R.J., Gruendl, R.A., Teuben, P.J. 1993, \pasp,
	105, 666
\reference{VB} von Hippel, T., \& Bothun, G. 1990, \aj, 100, 403


\end{references}
\end{document}